\documentclass[10pt,twocolumn]{article}
\usepackage{amssymb,amsmath,epsfig}

\begin{document}

\title{\bf Charge Gravastars in $f(T)$ Modified Gravity}

\author{\bf{Ujjal Debnath}\thanks{ujjaldebnath@gmail.com}\\
Department of Mathematics, Indian Institute of Engineering\\
Science and Technology, Shibpur, Howrah-711 103, India.\\\\\\}

\maketitle

\begin{abstract}
In the present work, we have discussed the four dimensional
spherically symmetric steller system in the framework of modified
$f(T)$ gravity theory with electro-magnetic field. The field
equations have been written for two cases, either $T'=0$ or
$f_{TT}=0$. Next we have discussed the charged gravastar model
which has three regions: interior region, shell region and
exterior region. In the interior region, we have found the
solutions of all physical quantities like density, pressure,
electro-magnetic field and also the metric coefficients for both
the cases. For $T'=0$, gravastar cannot form but it forms only for
the case $f_{TT}=0$. In the exterior region, we have obtained the
exterior solution for vacuum model. In the shell region, we have
assumed that the interior and exterior regions join together at a
place, so the intermediate region must be thin shell with the
approximation $h\ll 1$. Under this approximation, we have found
the analytical solutions. The proper length of the thin shell,
entropy and energy content inside the thin shell have been found
and they are directly proportional to the proper thickness of the
shell $\epsilon$ under the approximation ($\epsilon\ll 1$).
According to the Darmois-Israel formalism, we have studied the
matching between the surfaces of interior and exterior regions of
the gravastar. The energy density,  pressure, equation of state
parameter on the surface and mass of the thin shell have been
obtained.
\end{abstract}

\section{Introduction}

In 2001, Mazur and Mottola \cite{Mazur,Mazur1} have found a
solution for the gravitationally collapsing neutral system in the
concept of Bose-Einstein condensation to gravitational systems.
These describe as super compact, spherically symmetric and {\it
singularity free} objects, that can be considered to be virtually
as compact as the black holes. These gravitationally dark cold
vacuum compact star is known as {\it gravastar} (``gravitational
vacuum condensate stars"). The gravastar is a substitute of black
hole i.e., the existence of compact stars minus event horizons,
which compresses the matter within the gravitational radius
$R=\frac{2GM}{c^{2}}$, that is very close to the Schwarzschild
radius. In this sense, the existence of quantum vacuum
fluctuations are predicted near the event horizon. The gravastar
consists of an (i) interior de Sitter condensate phase and (ii)
exterior Schwarzschild geometry. The gravastar consist of five
layers where infinitely {\it two thin shells}, apparently, on the
regions $r = r_{1}$ and $r = r_{2}$ where $r_{1}$ and $r_{2}$ are
inner and outer radii ($r_{1}<r_{2}$). Also other important
regions are (i) {\it Interior region}: $0 \le r < r_{1}$ with
equation of state (EOS) $p=-\rho$ which is defined by the
de-Sitter spacetime, (ii) {\it Shell region}: $r_{1} < r < r_{2}$
with EOS $p=\rho$ i.e., an intermediary thin layer made of
ultra-stiff perfect fluid, (iii) {\it Exterior region}: $r_{2}<r$
with EOS $p =\rho= 0$ (vacuum) which is described by the
Schwarzschild solution. Thus the interior region of the gravastar
is surrounded by a thin shell of ultra-relativistic matter whereas
the exterior region is completely vacuum. Also, if we replace two
infinitely thin stiff shells and the intermediary region with just
one infinitely thin shell core \cite{Visser}, then the five layer
models can be simplified to the three layer models.

There are lot of works on the gravastar models available in the
literature in the framework of Einstein's general relativity.
DeBenedictis et al \cite{De} have found gravastar solutions with
continuous pressures and equation of state. Cattoen et al
\cite{Cattoen} have taken anisotropic pressure in the formation of
gravastar. Born-infeld phantom gravastars have been discussed by
Bilic et al \cite{Bilic}. Gravastar model in higher dimensional
spacetime has been discussed in refs
\cite{Farook1,Bhar,Ghos,Ghosh1}. Gravastar in the frame of
conformal motion has also been analyzed by some authors
\cite{Bhar,Usmani,Ayan}. Physical features of charged gravastar
have been investigated
\cite{Bhar,Ghos,Usmani,Horvat,Chan,Rahaman1,Brandt,Ali}. Stable
gravastar is discussed by several authors
\cite{Bene,Rocha1,Rocha2,Chan1,Ayan1}.

One important family of modifications of Einstein-Hilbert action
is the $f(R)$ theories of gravity \cite{Noj1,Capo,Soti,Feli}. In
such theories, one can uses a function of curvature scalar as the
Lagrangian density. In the similar line, one can also modify
teleparallel equivalent of general relativity where Lagrangian
density is equivalent to the torsion scalar $T$ and the field
equations of teleparallel gravity \cite{Ein,Ein1,Ein2,Unzi,And}
are identical with the Einstein's field equations in any
background metric. After that it has been modified teleparallel
gravity by having a Lagrangian density equivalent to a function of
torsion scalar i.e., $f(T)$ gravity \cite{Beng}. In theoretical
astrophysics, $f(T)$ version of 3-dimensions of BTZ black hole
solutions has been calculated as $f(T)$ gravity theory was
supported for examining the effects of $f(T)$ gravity models
\cite{Dent}. Recently, static solutions in the spherically
symmetric with charged source in $f(T)$ gravity have been found
\cite{Wang00}.

Deliduman et al \cite{Del} have investigated the structure of
neutron stars in modified $f(T)$ gravity models. Boehmer et al
\cite{Boem} have analyzed the existence of relativistic stars in
$f(T)$ modified gravity and explicitly constructed several classes
of static perfect fluid solution. Anisotropic strange quintessence
stars in $f(T)$ gravity models have been studied in refs
\cite{Abha1,Saha1}. Compact stars of emending class one in $f(T)$
gravity has been studied in \cite{Abha2}. Strong magnetic field
effects on neutron stars within $f(T)$ theory of gravity has been
discussed in \cite{Gan}. Tolman-Oppenheimer-Volkoff equations and
their implications for the structures of relativistic stars in
$f(T)$ gravity have been found \cite{Kp}.  Abbas and his
collaborations \cite{Abbas6,Abbas7,Abbas8,Abbas9} have discussed
the anisotropic compact star models in GR, $f(R)$, $f(G)$ and
$f(T)$ theories in diagonal tetrad case with Krori and Barua (KB)
metric \cite{Krori}. Gravastar model in $f(G,{\cal T})$ gravity
has been discussed by Shamir et al \cite{Shamir}. Das et al
\cite{Das1} have studied the gravastar model in modified
$f(R,{\cal T})$ gravity.

Main motivation of the work is to study the gravastar in the
framework of $f(T)$ gravity with the electromagnetic source and
examine the nature of physical parameters in the thin shell
region. The organization of the work is as follows: In section 2,
we have presented some brief reviews of $f(T)$ gravity with
electromagnetic field. In section 3, we have written
Einstein-Maxwell field equations for spherically symmetric steller
metric in the framework of $f(T)$ gravity theory. In section 4, we
have devoted the three regions of gravastar model and found the
metric co-efficients and other physical quantities. In section 5,
the physical features of gravastar parameters have been analyzed.
The junction conditions between interior and exterior regions have
been studied in section 6. Some physical analysis and conclusions
of the work have been written in section 7.

\section{$f(T)$ Modified Gravity and Electromagnetic Field}

We consider the torsion and the con-torsion tensor as follows
\cite{Abha1}:
\begin{equation}\label{1}
T_{\mu\nu}^{\alpha}=\Gamma_{\nu\mu}^{\alpha}-\Gamma_{\mu\nu}^{\alpha}=e_{i}^{\alpha}(\partial_{\mu}e_{\nu}^{i}-\partial_{\nu}e_{\mu}^{i})
\end{equation}

\begin{equation}\label{2}
K_{\alpha}^{\mu\nu}=-\frac{1}{2}(T_{\alpha}^{\mu\nu}-T_{\alpha}^{\nu\mu}-T_{\alpha}^{\mu\nu})
\end{equation}
and the components of the tensor $S_{\alpha}^{\mu\nu}$ are
defined as
\begin{equation}\label{3}
S_{\alpha}^{\mu\nu}=\frac{1}{2}(K_{\alpha}^{\mu\nu}+\delta_{\alpha}^{\mu}T_{\beta}^{\beta\nu}-\delta_{\alpha}^{\nu}T_{\beta}^{\beta\mu})
\end{equation}
So one can write the torsion scalar as in the following form:
\begin{equation}\label{4}
T=T_{\mu\nu}^{\alpha}S_{\alpha}^{\mu\nu}
\end{equation}
whose importance becomes clear in a moment. Now, define the
modified teleparallel action by replacing $T$ with a function
$f(T)$ \cite{Sotiriou,1BKCSNODS12} as follows

\begin{equation}\label{5}
S=\int d^{4}x\sqrt{-g}\left[\frac{1}{16\pi}f(T)+{\cal
L}_{Matter}\right]
\end{equation}
where we choose $G=c=1$ and ${\cal L}_{Matter}$ is the matter Lagrangian.\\

The ordinary matter is an anisotropic fluid so that the
energy-momentum tensor is given by

\begin{equation}\label{6}
T_{\mu\nu}^{matter}=(\rho+p)u_{\mu}u_{\nu}+pg_{\mu\nu}
\end{equation}
where $u^{\mu}$ is the fluid four-velocity satisfying
$u_{\mu}u^{\nu}=-1$, $\rho$ the energy density of fluid and $p$ is
the fluid pressure. Further, the energy momentum tensor for
electromagnetic field is given by

\begin{equation}\label{7}
T_{\mu\nu}^{EM}=-\frac{1}{4\pi}(g^{\delta\omega}F_{\mu\delta}F_{\omega\nu}-\frac{1}{4}g_{\mu\nu}F_{\delta\omega}F^{\delta\omega})
\end{equation}
where $F_{\mu\nu}$ is the Maxwell field tensor defined as

\begin{equation}\label{8}
F_{\mu\nu}=\Phi_{\nu,\mu}-\Phi_{\mu,\nu}
\end{equation}
and $\Phi_{\mu}$ is the four potential. The corresponding Maxwell
electromagnetic field equations are
\begin{equation}
(\sqrt{-g}~F^{\mu\nu}),_{\nu}=4\pi
J^{\mu}\sqrt{-g}~,~F_{[\mu\nu,\delta]}=0
\end{equation}
where $J^{\mu}$ is the current four-vector satisfying
$J^{\mu}=\sigma u^{\mu}$, the parameter $\sigma$ is the charge
density.

\section{Einstein-Maxwell Field equations}

We consider the spherically symmetric metric describing the
interior space-time as \cite{Krori}

\begin{equation}\label{9}
ds^{2}=-e^{a(r)}dt^{2}+e^{b(r)}dr^{2}+r^{2}(d\theta^{2}+\sin^{2}\theta
d\phi^{2})
\end{equation}
where $a(r)$ and $b(r)$ are functions of $r$. For this metric, we
get the torsion scalar $T$ and its derivative $T'$ as in the
following forms \cite{Saha1}:
\begin{equation}\label{11}
T(r)=\frac{2e^{-b}}{r}\left(a'+\frac{1}{r}\right)~,
\end{equation}
\begin{equation}\label{12}
T'(r)=\frac{2e^{-b}}{r}\left[a''-\frac{1}{r^{2}}-\left(a'+\frac{1}{r}\right)\left(b'+\frac{1}{r}\right)\right]
\end{equation}
where the prime $'$ denotes the derivative with respect to the
radial coordinate $r$.\\

For the charged fluid source with density $\rho(r)$, pressure
$p(r)$ and electromagnetic field $E(r)$, the Einstein-Maxwell (EM)
equations reduce to the form ($G=c=1$) \cite{Boem,Abha1}

\begin{eqnarray}\label{13}
-2\frac{e^{-b}}{r}T'f_{TT}+\frac{f}{2}-\left(T-\frac{1}{r^{2}}-\frac{e^{-b}}{r}(a'+b')\right)
f_{T} \nonumber\\
=8\pi\rho+E^{2}~,
\end{eqnarray}
\begin{equation}\label{14}
\left(T-\frac{1}{r^{2}}\right)f_{T}-\frac{f}{2}=8\pi p-E^{2}~,
\end{equation}
\begin{eqnarray*}
e^{-b}\left(\frac{a'}{2}+\frac{1}{r} \right)T'f_{TT}
+\left[\frac{T}{2}+e^{-b}\left\{\frac{a''}{2}+\left(\frac{a'}{4}+\frac{1}{2r}\right)\times
\right.\right.
\end{eqnarray*}
\begin{eqnarray}\label{15}
~~~~~~~~~~~~~~~~~~\left.\left. (a'-b')\right\}\right]
f_{T}-\frac{f}{2}=8\pi p+E^{2},
\end{eqnarray}
\begin{equation}\label{16}
\frac{e^{-\frac{b}{2}}\cot\theta}{2r^{2}}T'f_{TT}=0
\end{equation}

Adding equations (13) and (14), we obtain
\begin{equation}\label{17}
\rho+p=\frac{e^{-b}}{8\pi r} (a'+b')f_{T}
\end{equation}

Taking the derivative of equation (\ref{14}) and using the
equations (\ref{11}) and (\ref{17}), we obtain the energy
conservation equation \cite{Ghos}
\begin{equation}\label{18}
p'+\frac{1}{2}(\rho+p)a'=\frac{1}{8\pi r^{4}}(r^{4}E^{2})'
\end{equation}
and the electric field $E$ is as follows
\begin{equation}\label{19}
E(r)=\frac{1}{r^{2}}\int_{0}^{r}4\pi r^{2}\sigma(r)
e^{\frac{b}{2}}dr=\frac{q(r)}{r^{2}}
\end{equation}
where $q(r)$ is the total charge within a sphere of radius $r$.
The term $\sigma e^{\frac{b}{2}}$ inside the above integral is
known as the volume charge density. The gravitational mass can be
written as \cite{Ghos}
\begin{equation}
M(r)=\int_{0}^{r}4\pi r^{2}\left(\rho+\frac{E^{2}}{8\pi}\right) dr
\end{equation}

From equation (\ref{16}), we get either $T'=0$ or $f_{TT}=0$.\\

{\bf Case I:} $T'=0$~ $\Rightarrow~ T=$ constant $=T_{0}$, i.e.,
$T$ is independent of $r$ and hence $f(T),~f_{T},~f_{TT},....$ are
always constant. Assume $f(T_{0})=f_{0}$ and $f_{T}(T_{0})=f_{1}$.
So equations (\ref{13}) - (\ref{15}) reduce to
\begin{eqnarray}\label{21}
\frac{f_{0}}{2}-\left(T_{0}-\frac{1}{r^{2}}-\frac{e^{-b}}{r}(a'+b')\right)f_{1}
=8\pi\rho+E^{2}~,
\end{eqnarray}
\begin{equation}\label{22}
\left(T_{0}-\frac{1}{r^{2}}\right)f_{1}-\frac{f_{0}}{2}=8\pi
p-E^{2}~,
\end{equation}
\begin{eqnarray}\label{23}
\left[\frac{T_{0}}{2}+e^{-b}\left\{\frac{a''}{2}+\left(\frac{a'}{4}+\frac{1}{2r}\right)(a'-b')\right\}\right]
f_{1}-\frac{f_{0}}{2}\nonumber\\
=8\pi p+E^{2}
\end{eqnarray}

{\bf Case II:} $f_{TT}=0$~$ \Rightarrow~ f(T)=\alpha T+\beta$,
where $\alpha$ and $\beta$ are constants. Equations (\ref{13}) -
(\ref{15}) reduce to
\begin{equation}\label{24}
\frac{\alpha
e^{-b}}{r}\left(b'-\frac{1}{r}\right)+\frac{\alpha}{r^{2}}+\frac{\beta}{2}=8\pi\rho+E^{2}~,
\end{equation}
\begin{equation}\label{25}
\frac{\alpha
e^{-b}}{r}\left(a'+\frac{1}{r}\right)-\frac{\alpha}{r^{2}}-\frac{\beta}{2}=8\pi
p-E^{2}~,
\end{equation}
\begin{equation}\label{26}
\alpha e^{-b}\left[\frac{a''}{2}+\left(\frac{a'}{4}+\frac{1}{2r}
\right)(a'-b') \right]-\frac{\beta}{2}=8\pi p+E^{2}
\end{equation}

\section{Geometry of Gravastars}

In this section we will derive the solutions of the field
equations for gravastar and discuss its physical as well as
geometrical interpretations. For geometrical regions of the
gravastar, it is supposed to be extremely thin having a finite
width within the regions $D=r_{1}<r<r_{2}=D+\epsilon$ where
$r_{1}$ and $r_{2}$ are radii of the interior and exterior regions
of the gravastar and $\epsilon$ is positive small quantity. In
these regions, the equation of state (EOS) parameter is structured
as follows: (i) {\it Interior region} $R_{1}$: $0 \le r < r_{1}$
with EOS $p=-\rho$, (ii) {\it Shell region} $R_{2}$: $r_{1} < r <
r_{2}$ with EOS $p=\rho$, (iii) {\it Exterior region} $R_{3}$:
$r_{2}<r$ with EOS $p =\rho= 0$.

\subsection{Interior Region}

The equation of state for interior region $R_{1}$  ($0\le
r<r_{1}=D$) of the gravastar is $p=-\rho$. From equation
(\ref{19}), we obtain

\begin{equation}
e^{b(r)}=k~e^{-a(r)}
\end{equation}
where $k$ is constant $>0$.\\

$${\bf Case~I: ~~{\it T'=0}}$$

For $T'=0$, we obtain the solutions as
\begin{equation}
e^{a(r)}=\frac{kT_{0}r^{3}+C}{6r}
\end{equation}
and
\begin{equation}
e^{b(r)}=\frac{6kr}{kT_{0}r^{3}+C}
\end{equation}
where $C$ is constant. So in this case, the metric becomes
\begin{equation}
ds^{2}=-\frac{kT_{0}r^{3}+C}{6r}dt^{2}+\frac{6kr}{kT_{0}r^{3}+C}dr^{2}+r^{2}(d\theta^{2}+\sin^{2}\theta
d\phi^{2})
\end{equation}

Now we find the expressions of density, pressure and electric
field as
\begin{equation}
\rho=\frac{1}{16\pi}\left[f_{0}+\left(\frac{1}{r^{2}}-2T_{0}\right)f_{1}\right]~,
\end{equation}
\begin{equation}
p=\frac{1}{16\pi}\left[\left(2T_{0}-\frac{1}{r^{2}}\right)f_{1}-f_{0}\right]
\end{equation}
and
\begin{equation}
E(r)=\frac{\sqrt{f_{1}}}{\sqrt{2}~r}
\end{equation}
So the charge density for the electric field can be expressed as
\begin{equation}
\sigma=\sigma_{0}r^{m}\left(\frac{kT_{0}r^{3}+C}{6r}\right)^{\frac{1}{2}}
\end{equation}

Also the gravitational mass of the interior region can be found as
\begin{eqnarray}
M(D)=\int_{0}^{r_{1}=D}4\pi
r^{2}\left(\rho+\frac{E^{2}}{8\pi}\right)dr
\nonumber\\
=\frac{1}{12}(f_{0}-2f_{1}T_{0})D^{3}+\frac{1}{2}f_{1}D
\end{eqnarray}
Since gravastar is singularity free object, but we observe that
the central singularity always occurs at $r=0$. {\it So gravastar
cannot form in the case of $T'=0$}.

$${\bf Case~II: ~~{\it f_{TT}=0}}$$

For $f_{TT}=0$, there are 4 equations and 5 unknown functions
$a,~b,~\rho,~p,~E$. So one function is free. Let us assume $\sigma
e^{\frac{b}{2}}=\sigma_{0}r^{m}$ \cite{Ghos} where $\sigma_{0}$
and $m$ are constants, so from equation (\ref{18}) we have
\begin{equation}\label{35}
E(r)=E_{0}r^{m+1}
\end{equation}
where $E_{0}=\frac{4\pi \sigma_{0}}{m+3}$~. Using equation
(\ref{17}), we obtain
\begin{equation}
p=-\rho=k_{1}r^{2m+2}+k_{2}
\end{equation}
where $k_{1}=\frac{(m+3)E_{0}^{2}}{8\pi(m+1)}$ and $k_{2}$ is
constant. From equation (\ref{24}), we obtain
\begin{eqnarray}
e^{a}=ke^{-b}=k\left[1-\frac{k_{3}}{2\alpha r}+\frac{(\beta+16\pi
k_{2})}{6\alpha} ~r^{2} \right. \nonumber \\
\left. +\frac{(8\pi k_{1}-E_{0}^{2})}{\alpha(2m+5)}~r^{2m+4}
\right]
\end{eqnarray}
where $k_{3}$ is an integration constant. We see that for
$k_{3}\ne 0$, the central singularity occurs at $r=0$. Since
gravastar is singularity free object, so the metric will be
non-singular at the center $r=0$. Hence we can choose $k_{3}=0$.
So the metric becomes (choose $k=1$)
\begin{eqnarray}
ds^{2}=-\left[1+\frac{(\beta+16\pi k_{2})}{6\alpha}
~r^{2}+\frac{(8\pi k_{1}-E_{0}^{2})}{\alpha(2m+5)}~r^{2m+4}
\right]dt^{2}~~~~~~~~~~~~~~~~~~~~~~~~ \nonumber
\end{eqnarray}
\begin{eqnarray}
+\left[1+\frac{(\beta+16\pi k_{2})}{6\alpha} ~r^{2}+\frac{(8\pi
k_{1}-E_{0}^{2})}{\alpha(2m+5)}~r^{2m+4}
\right]^{-1}dr^{2} \nonumber\\
+r^{2}(d\theta^{2}+\sin^{2}\theta
d\phi^{2})
\end{eqnarray}
The charge density for electric field will be
\begin{equation}
\sigma=\sigma_{0}r^{m} \left[1+\frac{(\beta+16\pi k_{2})}{6\alpha}
~r^{2}+\frac{(8\pi k_{1}-E_{0}^{2})}{\alpha(2m+5)}~r^{2m+4}
\right]^{\frac{1}{2}}
\end{equation}

Also the gravitational mass of the interior region of the
gravastar can be found as
\begin{eqnarray}
M(D)=\int_{0}^{r_{1}=D}4\pi
r^{2}\left(\rho+\frac{E^{2}}{8\pi}\right) dr \nonumber\\
=\frac{(E_{0}^{2}-8\pi k_{1})}{2(2m+5)}~r^{2m+5}-\frac{4\pi
k_{2}}{3}~r^{3}
\end{eqnarray}
{\it Thus gravastar forms in the case of $f_{TT}=0$}. In the next
subsections, we'll consider only the case $f_{TT}=0$.

\subsection{Shell Region}
In this region $R_{2}$ ($D=r_{1}<r<r_{2}=D+\epsilon$), we assume
the thin shell contains stiff perfect fluid which obeys EoS
$p=\rho$. For this non-vacuum region, it is very difficult to
obtain the general solutions from the field equations. When two
region joins together at a place, the intermediate region must be
thin shell. So we shall try to find the analytical solution within
the thin shell with limit $0<e^{-b}\equiv h\ll 1$. This thin shell
structure suggests that as $r$ approaches to zero, the
corresponding radial parameter
generally becomes $\ll 1$. \\

For $f_{TT}=0$ and under the above approximation (we set $h$ be
zero to the leading order), the field equations (\ref{24}) -
(\ref{26}) reduce to the following forms \cite{Ghos}
\begin{equation}\label{45}
-\frac{\alpha h'}{r}+\frac{2\alpha}{r^{2}}+\beta=2E^{2}
\end{equation}
and
\begin{equation}\label{46}
\alpha
h^{'}\left(\frac{a'}{4}+\frac{1}{2r}\right)+\frac{\alpha}{r^{2}}=2E^{2}
\end{equation}
We see that there are two equations but three unknowns $a,~h$ and
$E$. Similar to equation (\ref{35}), let us assume the solution of
$E$ in the form $E(r)=E_{0}r^{m+1}$. Solving equations (\ref{45})
and (\ref{46}), we obtain
\begin{equation}\label{55}
e^{-b(r)}\equiv h(r)=h_{2}+2r+\frac{\beta
r^{2}}{2\alpha}-\frac{E_{0}^{2}}{\alpha(m+2)}~r^{2m+4}
\end{equation}
and
\begin{equation}
e^{a(r)}=\frac{h_{3}}{r^{2}}~Exp\left[\int\frac{8E_{0}^{2}r^{2m+4}-4\alpha
}{2\alpha r+\beta r^{3}-2E_{0}^{2}r^{2m+5}} ~dr\right]
\end{equation}
where $h_{2}$ and $h_{3}$ are integration constants and the radius
$r$ corresponds to the shell structure in the region $R_{2}$. In
this shell region, the range of $r$ is $D< r< D+\epsilon$. Under
this assumption ($h\ll 1$), $\epsilon\ll 1$, so we must have
$h_{2}\ll 1$. From equation (\ref{25}) we obtain
\begin{equation}
8\pi p=8\pi
\rho=E_{0}^{2}r^{2m+2}-\frac{\alpha}{r^{2}}-\frac{\beta}{2}
\end{equation}
Also the charge density for electric field is given by
\begin{equation}
\sigma=\sigma_{0}r^{m} \left[h_{2}+2r+\frac{\beta
r^{2}}{2\alpha}-\frac{E_{0}^{2}}{\alpha(m+2)}~r^{2m+4}
\right]^{\frac{1}{2}}
\end{equation}

\subsection{Exterior Region}
For the exterior region $R_{3}$ ($r>r_{2}=D+\epsilon$), the vacuum
EoS is given by ($p=\rho=0$). In this region, from equation
(\ref{17}), we obtain
\begin{equation}
E(r)=\frac{Q}{r^{2}}
\end{equation}
where $Q$ is constant electric charge. From equation (\ref{19}),
we obtain
\begin{equation}
e^{b(r)}=k~e^{-a(r)}
\end{equation}
where $k$ is constant $>0$.\\

For $f_{TT}=0$, we obtain the solutions as
\begin{equation}
e^{a}=ke^{-b}=k\left(1-\frac{2M}{
r}+\frac{\beta}{6\alpha}~r^{2}+\frac{Q^{2}}{\alpha r^{2}}\right)
\end{equation}
where $M$ is the mass of the gravastar. Also the charge density
for electric field may be written as
\begin{equation}
\sigma=\sigma_{0}r^{m} \left(1-\frac{2M}{
r}+\frac{\beta}{6\alpha}~r^{2}+\frac{Q^{2}}{\alpha
r^{2}}\right)^{\frac{1}{2}}
\end{equation}

So in the exterior region, the metric becomes (choose $k=1$)
\begin{eqnarray}
ds^{2}=-\left(1-\frac{2M}{
r}+\frac{\beta}{6\alpha}~r^{2}+\frac{Q^{2}}{\alpha
r^{2}}\right)dt^{2}\nonumber\\
+\left(1-\frac{2M}{
r}+\frac{\beta}{6\alpha}~r^{2}+\frac{Q^{2}}{\alpha
r^{2}}\right)^{-1}dr^{2}+r^{2}(d\theta^{2}+\sin^{2}\theta
d\phi^{2})
\end{eqnarray}
For $\alpha=1$ and $\beta=0$, we get usual Reissner-Nordstrom
spacetime metric and also for $\alpha=1$, $\beta=0$ and $Q=0$,
we get back to static Schwarzschild metric.\\

\section{Physical Features}
Now we shall discuss the physical features of the parameters of
the gravastar shell region like proper length of the shell, energy
and entropy within the shell.

\subsection{Proper Length}

Since the stiff perfect fluid propagates between two boundaries of
the shell region of the gravastar, so the inner boundary of the
shell is located at the surface $r=D$ and outer boundary of the
shell is located at the surface $r=D+\epsilon$, where the proper
thickness of the shell is assumed to be very small, i.e.,
$\epsilon\ll 1$. So, the proper thickness of the shell is
determined by \cite{Ghos}
\begin{equation}
\ell=\int_{D}^{D+\epsilon}\sqrt{e^{b(r)}}~dr
\end{equation}
Since in the shell region, the expressions of $e^{b(r)}$ is
lengthy, so we cannot found the analytical form of the above
integral. So let us assume $\sqrt{e^{b(r)}}=\frac{dg(r)}{dr}$, so
from the above integral we can get
\begin{equation}
\ell=\int_{D}^{D+\epsilon}\frac{dg(r)}{dr}~dr=g(D+\epsilon)-g(D)
\approx \left.\epsilon
\frac{dg(r)}{dr}\right|_{D}=\epsilon\sqrt{e^{b(D)}}
\end{equation}
where we have taken only the first order term of $\epsilon$ (since
$\epsilon\ll 1$, so $O(\epsilon^{2})\approx 0$). For $f_{TT}=0$,
the proper length will be (from equation (\ref{55}))
\begin{equation}
\ell=\epsilon \left[h_{2}-2D-\frac{\beta
D}{2}+\frac{E_{0}^{2}}{\alpha(m+2)}~D^{2m+4}
\right]^{-\frac{1}{2}}
\end{equation}
which implies proper length of the shell is proportional to the
thickness ($\epsilon$) of the shell.

\subsection{Energy}

The energy within the shell region of the gravastar is \cite{Ghos}
\begin{equation}
{\cal E}=\int_{D}^{D+\epsilon}4\pi
r^{2}\left[\rho+\frac{E^{2}}{8\pi}\right]~dr
\end{equation}

For $f_{TT}=0$, we can obtain the energy as
\begin{eqnarray}
{\cal E}=\int_{D}^{D+\epsilon}
\left[E_{0}^{2}r^{2m+4}-\frac{\alpha}{2}-\frac{\beta r^{2}}{4}
\right]~dr \nonumber \\
=\frac{E_{0}^{2}}{2m+4}\left[(D+\epsilon)^{2m+4}-D^{2m+4} \right]
\nonumber\\
-\frac{\epsilon\alpha}{2}-\frac{\beta}{12}\left[(D+\epsilon)^{3}-D^{3}
\right]\\
\approx \epsilon E_{0}^{2}D^{2m+4}-\frac{\epsilon
\alpha}{2}-\frac{\epsilon\beta D^{2}}{4}~~~~~~~~
\end{eqnarray}
For this approximation, we see that the energy content in the
shell is proportional to the thickness ($\epsilon$) of the shell.

\subsection{Entropy}
Mazur and Mottola \cite{Mazur,Mazur1} have shown that the entropy
density is zero in the interior region $R_{1}$ of the gravastar.
However, the entropy within the shell can be defined by
\cite{Ghos}
\begin{equation}
S=\int_{D}^{D+\epsilon} 4\pi r^{2} s(r) \sqrt{e^{b(r)}}~dr
\end{equation}
where $s(r)$ is the entropy density for the local temperature
$T(r)$ and which can be written as
\begin{equation}
s(r)=\frac{\gamma^{2}k_{B}^{2}T(r)}{4\pi \hslash}=\frac{\gamma
k_{B}}{\hslash}\sqrt{\frac{p(r)}{2\pi}}
\end{equation}
where $\gamma$ is dimensionless constant. So entropy can be
written as
\begin{equation}
S=\frac{\gamma k_{B}}{\hslash}\int_{D}^{D+\epsilon} r^{2}
\sqrt{8\pi p(r)~e^{b(r)}}~dr
\end{equation}
For the approximation ($\epsilon\ll 1$), we can obtain
\begin{equation}
S\approx\frac{\epsilon\gamma k_{B}}{\hslash}~D^{2} \sqrt{8\pi
p(D)~e^{b(D)}}
\end{equation}
For $f_{TT}=0$, we get the entropy as
\begin{eqnarray}
S\approx\frac{\epsilon\gamma
k_{B}}{\hslash}~\left[E_{0}^{2}D^{2m+3}-\frac{\alpha}{D}-\frac{\beta
D}{2}\right]^{\frac{1}{2}}\times \nonumber\\
\left[h_{2}-2D-\frac{\beta
D}{2}+\frac{E_{0}^{2}}{\alpha(m+2)}~D^{2m+4}
\right]^{-\frac{1}{2}}
\end{eqnarray}
That means entropy in the shell is proportional to the thickness
($\epsilon$) of the shell.

\section{Junction Conditions between Interior and Exterior Regions}

Since gravastar consist of three regions: interior region, shell
region and exterior region, so it is necessary to matching between
the surfaces interior and exterior regions according to the
Darmois-Israel formalism \cite{Is1,Is2,Is3}. At $r=D$, the
junction surface is denoted by $\Sigma$. We consider the metric on
the junction surface as in the form
\begin{equation}
ds^{2}=-f(r)dt^{2}+\frac{dr^{2}}{f(r)}+r^{2}(d\theta^{2}+\sin^{2}\theta
d\phi^{2})
\end{equation}
where the metric co-efficients are continuous at $\Sigma$, though
their derivatives may not be continuous at $\Sigma$. With the help
of Darmois-Israel formalism, we want to find the expression for
the stress-energy surface $S_{ij}$ from the Lanczos equation
\cite{Lanc}
\begin{equation}
S_{j}^{i}=-\frac{1}{8\pi}(\eta_{j}^{i}-\delta_{j}^{i}\eta_{k}^{k})
\end{equation}
where $\eta_{ij}=K_{ij}^{+}-K_{ij}^{-}$. Here $K_{ij}$ is the
extrinsic curvature. So $\eta_{ij}$ gives the discontinuous
surfaces in the extrinsic curvatures (second fundamental forms).
Here the signs $``+"$ and $``-"$ correspond to the interior and
the exterior regions of the gravastar respectively. The extrinsic
curvatures associated with the both surfaces of the shell region
can be written as
\begin{equation}
K_{ij}^{\pm}=\left[-n_{\nu}^{\pm}\left\{\frac{\partial^{2}x_{\nu}}{\partial
\xi^{i}\partial \xi^{j}}+\Gamma_{\alpha\beta}^{\nu}\frac{\partial
x^{\alpha}}{\partial \xi^{i}}\frac{\partial x^{\beta}}{\partial
\xi^{j}} \right\}\right]_{\Sigma}
\end{equation}
where $\xi^{i}$ are the intrinsic coordinates on the shell,
$n_{\nu}^{\pm}$ are the unit normals to the surface $\Sigma$,
defined by $n_{\nu}n^{\nu}=-1$. For the above metric, we can
obtain
\begin{equation}
n_{\nu}^{\pm}=\pm\left[g^{\alpha\beta}   \frac{\partial
f}{\partial x^{\alpha}} \frac{\partial f}{\partial
x^{\beta}}\right]^{-\frac{1}{2}} \frac{\partial f}{\partial
x^{\nu}}
\end{equation}
Using the Lanczos equation, we can obtain the stress-energy
surface tensor as $S^{i}_{j}=diag(-\varrho,\wp,\wp,\wp)$ where
$\varrho$ is the surface energy density and $\wp$ is the surface
pressure given by \cite{Ghosh1}
\begin{equation}
\varrho=-\frac{1}{4\pi D}\left[\sqrt{f} \right]_{-}^{+}
\end{equation}
and
\begin{equation}
\wp=\frac{1}{8\pi D}\left[\sqrt{f}
\right]_{-}^{+}+\frac{1}{16\pi}\left[\frac{f'}{\sqrt{f}}
\right]_{-}^{+}
\end{equation}

For $f_{TT}=0$, we obtain
\begin{eqnarray}
\varrho=-\frac{1}{4\pi D}\left[\sqrt{1-\frac{2M}{D}+\frac{\beta
D^{2}}{6\alpha}+\frac{Q^{2}}{\alpha D^{2}}} \right.
\nonumber\\
\left. -\sqrt{1+\frac{(\beta+16\pi
k_{2})D^{2}}{6\alpha}+\frac{(8\pi
k_{1}-E_{0}^{2})}{\alpha(2m+5)}~D^{2m+4}} \right]
\end{eqnarray}
and
\begin{eqnarray}
\wp=\frac{1}{8\pi D}\left[\sqrt{1-\frac{2M}{ D}+\frac{\beta
D^{2}}{6\alpha}+\frac{Q^{2}}{\alpha
D^{2}}} \right.\nonumber\\
\left.-\sqrt{1+\frac{(\beta+16\pi
k_{2})D^{2}}{6\alpha}+\frac{(8\pi
k_{1}-E_{0}^{2})}{\alpha(2m+5)}~D^{2m+4}} \right] \nonumber \\
+\frac{1}{16\pi}\left[\frac{\frac{2M}{D^{2}}+\frac{\beta
D}{3\alpha}-\frac{2Q^{2}}{\alpha D^{3}} }{\sqrt{1-\frac{2M}{
D}+\frac{\beta D^{2}}{6\alpha}+\frac{Q^{2}}{\alpha D^{2}}}}
\right.
\nonumber\\
\left.-\frac{\frac{(\beta+16\pi
k_{2})D}{3\alpha}+\frac{(2m+4)(8\pi
k_{1}-E_{0}^{2})}{\alpha(2m+5)}~D^{2m+3}
}{\sqrt{1+\frac{(\beta+16\pi k_{2})D^{2}}{6\alpha}+\frac{(8\pi
k_{1}-E_{0}^{2})}{\alpha(2m+5)}~D^{2m+4}}} \right]~~~~~~~~~~~
\end{eqnarray}

\subsection{Equation of State}
The equation of state parameter $w(D)$ can be written as
\begin{equation}
w(D)=\frac{\wp}{\varrho}
\end{equation}

For $f_{TT}=0$, the equation of state parameter can be written in
the following form
\begin{eqnarray}
w(D)=-\frac{1}{2}-\frac{1}{4}
\left[\frac{\frac{2M}{D^{2}}+\frac{\beta
D}{3\alpha}-\frac{2Q^{2}}{\alpha D^{3}} }{\sqrt{1-\frac{2M}{
D}+\frac{\beta D^{2}}{6\alpha}+\frac{Q^{2}}{\alpha D^{2}}}}
\right.
\nonumber\\
\left. -\frac{\frac{(\beta+16\pi
k_{2})D}{3\alpha}+\frac{(2m+4)(8\pi
k_{1}-E_{0}^{2})}{\alpha(2m+5)}~D^{2m+3}
}{\sqrt{1+\frac{(\beta+16\pi k_{2})D^{2}}{6\alpha}+\frac{(8\pi
k_{1}-E_{0}^{2})}{\alpha(2m+5)}~D^{2m+4}}} \right]
\nonumber\\
\times\left[\sqrt{1-\frac{2M}{D}+\frac{\beta
D^{2}}{6\alpha}+\frac{Q^{2}}{\alpha D^{2}}} \right. \nonumber\\
\left. -\sqrt{1+\frac{(\beta+16\pi
k_{2})D^{2}}{6\alpha}+\frac{(8\pi
k_{1}-E_{0}^{2})}{\alpha(2m+5)}~D^{2m+4}} \right]^{-1}
\end{eqnarray}

\subsection{Mass}

The mass ${\cal M}$ of the thin shell can be obtained from the
following formula
\begin{equation}
{\cal M}=4\pi D^{2}\varrho
\end{equation}

For $f_{TT}=0$, the mass of the thin shell can be expressed as
\begin{eqnarray}
{\cal M}=-D\left[\sqrt{1-\frac{2M}{D}+\frac{\beta
D^{2}}{6\alpha}+\frac{Q^{2}}{\alpha D^{2}}} \right.
\nonumber~~~~~~~~~~~~~~~~\\
\left. -\sqrt{1+\frac{(\beta+16\pi
k_{2})D^{2}}{6\alpha}+\frac{(8\pi
k_{1}-E_{0}^{2})}{\alpha(2m+5)}~D^{2m+4}} \right]
\end{eqnarray}
So the total mass $M$ of the gravastar in term of the thin shell
can be expressed as
\begin{eqnarray}
M=\frac{D}{2}+\frac{\beta D^{3}}{12\alpha}+\frac{Q^{2}}{2\alpha
D}~~~~~~~~~~~~~~~~~~~~~~~~~~~~~~~~~~~~~~~~~~~~~
\nonumber\\
-\frac{D}{2}\left[\sqrt{1+\frac{(\beta+16\pi
k_{2})D^{2}}{6\alpha}+\frac{(8\pi
k_{1}-E_{0}^{2})}{\alpha(2m+5)}~D^{2m+4}}~-\frac{{\cal M}}{D}
\right]^{2}
\end{eqnarray}
We see that the total mass $M$ of the gravastar will be less than
$\frac{D}{2}+\frac{\beta D^{3}}{12\alpha}+\frac{Q^{2}}{2\alpha
D}$.

\section{Discussions}

In the present work, we have discussed the four dimensional
spherically symmetric steller system in the framework of modified
$f(T)$ gravity theory with electro-magnetic field. The field
equations have been found for two cases, either $T'=0$ or
$f_{TT}=0$. For $T'=0$, $T$ must be constant and all the
derivatives of $f(T)$ with respect to $T$ must be constants. Also
for $f_{TT}=0$, we have obtained $f(T)$ is a linear function of
$T$. Next we have discussed the charged gravastar model where the
equation of state in the three regions of the gravastar satisfies
as follows: interior region ($p=-\rho$), shell region ($p=\rho$)
and exterior region ($p=\rho=0$). In the interior region, we have
found the solutions of all physical quantities like density,
pressure, electro-magnetic field and also metric coefficients for
both cases. For $T'=0$, we have found $E(r)\propto \frac{1}{r}$.
Since for $T'=0$, the central singularity occurs at $r=0$, so
gravastar cannot form in this case. For $f_{TT}=0$, $E(r)\propto
r^{m+1}$ and gravastar forms in this case. In the exterior region,
we have obtained the exterior solution for vacuum model. For
$f_{TT}=0$, we found that the metric is generalization of
Reissner-Nordstrom spacetime. In the shell region, we have assumed
that the interior and exterior regions join together at a place,
so the intermediate region must be thin shell with limit
$0<e^{-b}\equiv h\ll 1$. This thin shell structure suggests that
as $r$ approaches to zero, the corresponding radial parameter
generally becomes $\ll 1$. Under this approximation, we have found
the solutions for $f_{TT}=0$. The electromagnetic field becomes in
the form $E\propto r^{m+1}$. The proper length of the thin shell,
entropy and energy content inside the thin shell have been found
and they are directly proportional to the proper thickness of the
shell $\epsilon$ under the approximation ($\epsilon\ll 1$).
According to the Darmois-Israel formalism, we have studied the
matching between the surfaces interior and exterior regions of the
gravastar. The energy density and pressure on the surface have
been obtained. Also the equation of state parameter $w(D)$ have
been found. Moreover, the mass ${\cal M}$ of the thin shell have
been obtained and the total mass of the gravastar have been
expressed in terms of the thin shell mass.

\end{document}